# Dielectric properties of a ferroelectric nematic material: quantitative test of the polarization-capacitance Goldstone mode

Alex Adaka[1,3], Mojtaba Rajabi[2], Nilanthi Haputhantrige[2,3], S. Sprunt[2,3], O.D. Lavrentovich[1,2,3] and A. Jákli[,2,3]

[1]Materials Science Graduate Program, Kent State University, Kent OH, 44242, USA

[2]Department of Physics, Kent State University, Kent OH, 44242, USA

[3]Advanced Materials and Liquid Crystal Institute, Kent State University, Kent OH, 44242, USA

**Abstract**

The recently discovered ferroelectric nematic ($N_F$) liquid crystals (LC) have been reported to show an extraordinarily large value of the real part of the dielectric constant ($\varepsilon'$ >$10^3$) at low frequencies. However, it was argued by Clark et al in Physical Review Research 6, 013195 (2024) that what was measured was the capacitance of the insulating layer at LC/electrode surface and not that of the liquid crystal. Here we describe the results of dielectric spectroscopy measurements of an $N_F$ material in cells with variable thickness of the insulating layers. Our measurements quantitatively verify the model by Clark et al. Additionally, our measurements in cells with bare conducting indium tin oxide surface provide a crude estimate of $\varepsilon_\perp \sim 10^2$ in the $N_F$ phase.

The recently discovered ferroelectric nematic ($N_F$) liquid crystals [1–5] exhibit large ferroelectric polarization $P$ in the range of $(4-7) \times 10^{-2}\ C/m^2$ that is unprecedented in liquid crystals, and are also reported to show an extraordinarily large real part of the dielectric permittivity ($\varepsilon'$ >$10^3$) at low frequencies [3,4,6–9]. Recently, Clark et al. [10,11] have argued that the measured large dielectric constant of $N_F$ is an artifact and what has been measured is the capacitance of the non-ferroelectric interfacial insulating layer at the $N_F$/electrode surface and not that of the liquid crystal. This is due to the block-reorientation of the large polarization **P** of $N_F$, which renders the liquid crystal layer conductive, enabling the insulating interfacial layers to charge up. This mechanism leads to a polarization-capacitance Goldstone (PCG) mode with the apparent real $\varepsilon'_A(\omega)$ and imaginary $\varepsilon''_A(\omega)$ components of the dielectric spectra [10,11],

$$\varepsilon'_A(\omega) = \varepsilon_A(0)\frac{1}{1+(\omega\tau_o)^2} \text{ and } \varepsilon''_A(\omega) = \varepsilon_A(0)\frac{\omega\tau_o}{1+(\omega\tau_o)^2}. \qquad (1)$$

Here $\varepsilon_A(0) = \varepsilon_I d/d_I$, where $\varepsilon_I$ and $d_I$ are the dielectric constant and thickness of the insulating layer, respectively, and $d$ is the thickness of the liquid crystal film. Furthermore, the characteristic relaxation time in the $N_F$ cell is $\tau_o = \frac{\varrho_{LC}\varepsilon_o\varepsilon_I d}{d_I} = R_{LC}C_I$, where $\varrho_{LC} = \gamma/P^2$ is the effective resistivity of the LC, $\gamma$ is the rotational viscosity, and $C_I$ is the capacitance of the insulating layer. This results in an approximation for the relaxation frequency $f_o^{N_F}$ as

$$f_o^{N_F} = \frac{1}{2\pi\tau_o} = \frac{P^2}{2\pi\gamma\varepsilon_o\varepsilon_I} \cdot \frac{d_I}{d} \qquad (2)$$

More recently, Vaupotic et al. [12] studied homologues series of RM734 [1,4,13] sandwiched between conducting layers. The dielectric constant in the $N_F$ phase was found to exceed $10^4$, in accordance with previous measurements [3,4,6–8]. To interpret the results, Vaupotic et al. [12] proposed a continuous phenomenological model (CPM) in which the dielectric response is dominated by flexoelectricity, being relevant only at a very weak surface anchoring. In this condition the relaxation frequency $f_o$ becomes thickness dependent, $f_o \propto \left(\alpha + \frac{\beta}{d} + \frac{c}{d^2}\right)$, where $\alpha, \beta$ and $c$ are phenomenological constants.

Experimentally Erkoreka et al found that the measured dielectric strength (relaxation frequency) is proportional (inversely proportional) to the film thickness at low frequency, and independent of film thickness at high frequency. [14,15] This behavior is in qualitative agreement with the PCG [11], the CPM [12] (for $\alpha = c = 0$), and also with the electrode polarization (EP) mode known for conductive samples with electric double layers near the substrates [16]. In fact, the PCG model can be considered a particular realization of the EP mode, as the ferroelectric nematic can be considered as the "conductive sample" and $C_I$ relates to "electric double layers" near the substrates.

Finally, very recently Matko et al [17] analyzed impedance measurements carried out on several RM734 cells of different thicknesses $d$ and based on their simulation results they concluded "that the relative permittivity of the ferroelectric nematics is indeed huge, and it is even higher than the apparent measured values".

In this paper we describe the results of dielectric spectroscopy measurements of an $N_F$ material in cells with a controlled thickness of the insulating alignment layer (as was done for SmC* materials by Coleman et al [18]) and cells with bare conducting indium tin oxide (ITO) surface. We find excellent quantitative agreement with Eq.(1) and good agreement with Eq.(2) derived by Clark et al using the PCG model [11].

We study a liquid crystal UUZU-4-N provided by Merck [19]. On cooling, the material exhibits three nematic phases labeled as N, $N_X$, and $N_F$ with phase sequence in °C: I $-$ 94.1 $-$ N $-$ 92.4 $-$ $N_X$ $-$ 86.2 $-$ $N_F$ $-$ 60 $-$ Cr. The molecular structure of UUZU-4-N together with the temperature dependence of the ferroelectric polarization and the switching voltages are shown in Figure S1 of the Supplemental Material [20] The calculated (B3LYP/6-31G(d)) dipole moment of a single molecule is $\mu \approx 11.9\ Debye$ [19]. Results of the switching time measurements are described and shown in Figure S2 in the Supplemental Material [20] which includes [21–23]. Details of sample preparation, including the technique of preparing insulating PI2555 polyimide insulating layers with controlled thicknesses $d_I/2$, and the dielectric measurement technique are also described in the Supplemental Material [20].

The frequency dependences of the imaginary part of the dielectric constant $\varepsilon''$ at several selected temperatures for four cells with different insulating layer thickness $d_I$ values are shown in Figure 1(a-d).

In the $N_F$ phase (below $86°C$) the apparent dielectric spectra $\varepsilon_A(f)$ show a peak at frequencies corresponding to the relaxation frequencies $f_o^{N_F}$. Assuming Debye relaxation, the maxima correspond to half of the susceptibility $\frac{\chi}{2} = \frac{\varepsilon(0)-\varepsilon(\infty)}{2}$. It is found that $f_o^{N_F}$ increases from ~15 $kHz$ to 150 $kHz$, and $\chi$ decreases from ~360 to ~100 when $d_I$ increases from ~90 $nm$ to ~334 $nm$.

In the N phase the relaxation frequencies vary from a few hundred $Hz$ to a few $kHz$, i.e., an order of magnitude smaller than in the $N_F$ phase, while their $d_I/d$ dependences show similar trends to that in the $N_F$ phase. In the $N_X$ phase there are two peaks; one in the $kHz$, and one in the $10 - 100\ kHz$ range. The closer to the $N_F$ phase, the larger (smaller) is the maxima of the higher (lower) frequency peaks, indicating that the $N_X$ phase has coexisting $N_F$ and N domains. Optically this coexistence cannot be seen, which indicates that the size of these domains is below 1 µ$m$. In the I, $N_X$ and N phases we see an additional peak above 300 $kHz$.

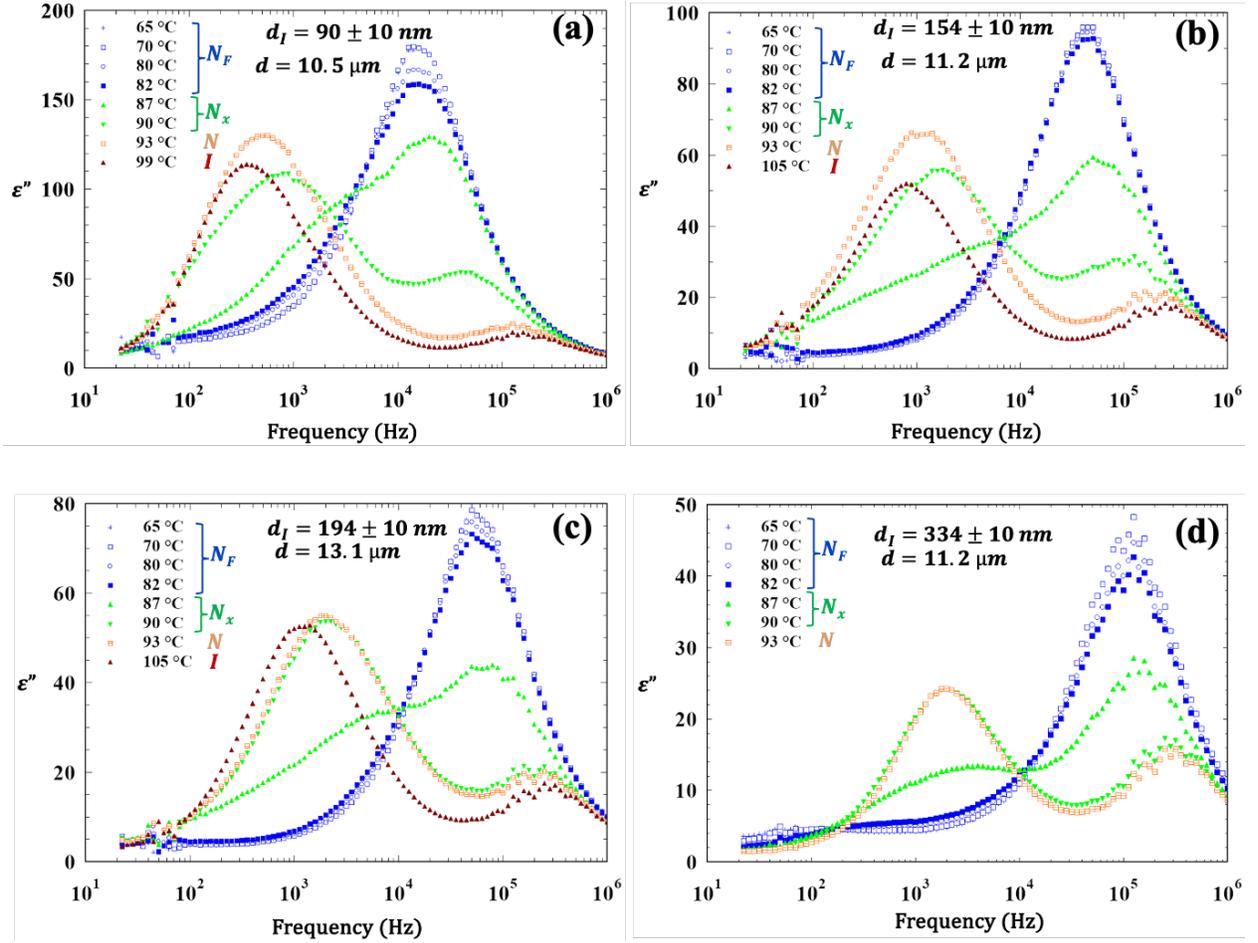

Figure 1: Frequency dependences of the imaginary part of the dielectric constant ε" at several selected temperatures for (a): d ≈ 10.5 µm and $d_I$ ≈ 90 nm; (b): d ≈ 11.2 µm and $d_I$ ≈ 154 nm; (c): d ≈ 13.1 µm and $d_I$ ≈ 194 nm; (d) d ≈ 11.2 µm and $d_I$ ≈ 334 nm film and alignment layer thicknesses.

To verify Eq.(1), in Figure 2 (a-c) we plot the measured $\varepsilon'$ values versus $d/d_I$ at 100 $Hz$, 1 $kHz$, and at 10 $kHz$. 100 $Hz$ is well below the relaxation frequencies, thus $\varepsilon'$ is approximately equal to $\varepsilon'_A(0)$; 1 $kHz$ is close to the relaxation frequencies found in the $N_X$ and N phases; and 10 $kHz$ is below the relaxation frequencies found in the $N_F$ phase, but above the relaxation frequencies found in the $N_X$ and N phases. According to Eq.(1), the slope of $\varepsilon'_A(d/d_I)$ should give $\frac{\varepsilon_I}{1+(\omega\tau_o)^2}$. It can be seen in Figure 2 (a-c) that within the measurement error, the $d/d_I$ dependence of $\varepsilon'_A$ is linear at all temperatures. Notably, at $100\ Hz$, in the $N_F$ phase the slopes of the best fits are nearly constant (~3.5), corresponding to the dielectric constant of the polyimide PI2555 [24]. At $1\ kHz$ (see Figure 2b) and at $10\ kHz$ (see Figure 2c) the slopes in the $N_F$ phase are still almost temperature independent, but they have smaller values: ~3.3 for $1\ kHz$ and ~2.5 at $10\ kHz$, a decrease caused by the factor $\frac{1}{1+(\omega\tau_o)^2}$. These observations quantitatively agree with Eq.(1), and thus completely support the model by Clark et al [11].

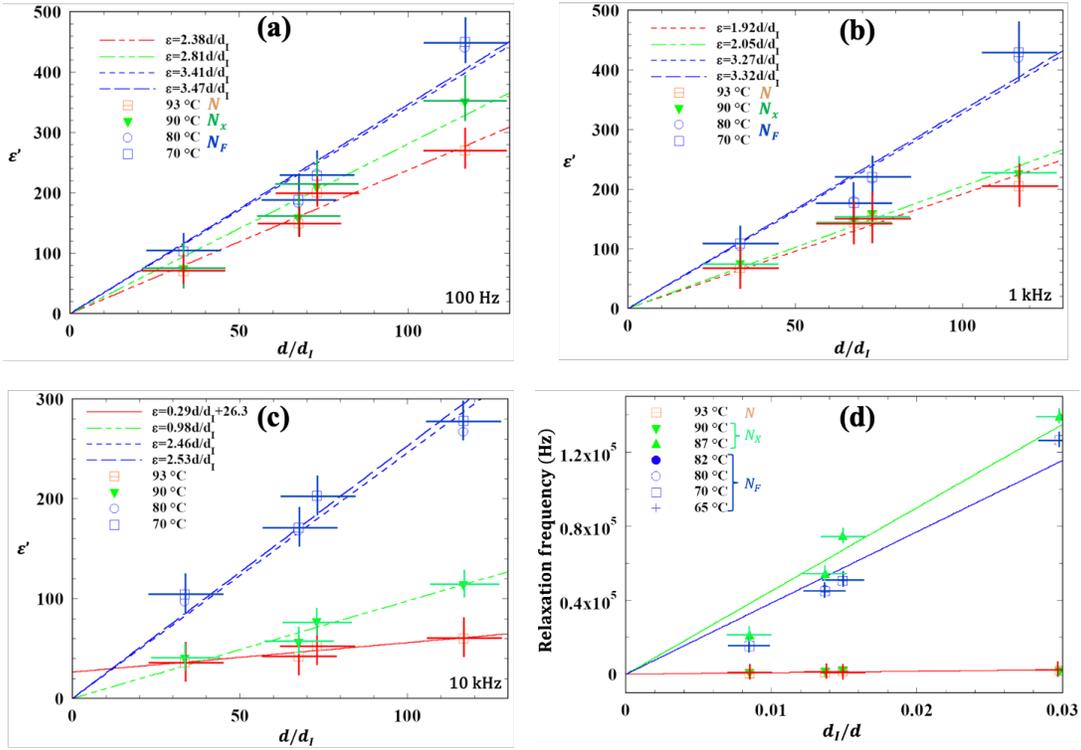

*Figure 2: The measured $\varepsilon'$ as a function of $d/d_I$ at $100\ Hz$ (a), $1\ kHz$ (b), and at $10\ kHz$ (c) at various temperatures. (d) The relaxation frequency as a function of $d_I/d$ at various temperatures. Lines represent the best fits with slopes $8 \cdot 10^4\ Hz$, $4.5 \cdot 10^6\ Hz$ and $3.85 \cdot 10^6\ Hz$ in the $N, N_X$ and $N_F$ phases, respectively.*

Although the PCG model is not applicable to the N phase, we find a similar $d/d_I$ dependence with a much smaller slope than in the $N_F$ phase. This behavior is likely related to the large conductivity with electric double layer near the substrate (EP mode) [16]. The decrease of the slopes in the $N_X$ phase is in between those in the $N_F$ and N phases, which is understandable, taking into account that $N_X$ is characterized both by the low and high frequency relaxations.

The $d_I/d$ dependence of the measured relaxation frequencies at several temperatures in the $N_F$, $N_X$ and N phases are shown in Figure 2d. The best fits assuming $f_o \propto \frac{d_I}{d}$ corresponding to Eq.(2) have almost the same slopes (in the range of $(3.8 - 4.4) \cdot 10^6 s^{-1}$) in the $N_F$ phase, $\sim 4.5 \cdot 10^6 s^{-1}$ the $N_X$ phase, and $8 \cdot 10^4 s^{-1}$ in the $N$ phase. In the $N_F$ phase Eq.(2) gives $\gamma \approx \frac{(6 \cdot 10^{-2})^2}{2\pi \cdot (3.8-4.4) \cdot 10^6 \cdot 3.1 \cdot 10^{-11}} \sim (4.2 - 4.9) \, Pa \cdot s$. These values are several times larger than what we obtained from the switching time measurements (see Supplemental Material and Figure S2). This difference is not surprising, since the dielectric experiments in sandwich cell mainly test splay deformation, while the in-plane field-induced, full polarization switching involves twist.

In the N and I phases the polarization is zero, and Eq.(2) is not applicable. We propose that the measured large dielectric constant below the $kHz$ relaxation frequency range is due to the high conductivity of UUZU-4-N in the isotropic phase. Following the argument of Clark et al [11], $R_{LC} = \frac{d}{A\sigma}$, so the relaxation time is $\tau_o = R_{LC} \cdot C_I = \frac{\varepsilon_o \varepsilon_I}{\sigma} \cdot \frac{d}{d_I}$. Accordingly,

$$f_o^N = \frac{1}{2\pi\tau_o} = \frac{\sigma}{2\pi\varepsilon_o\varepsilon_I} \cdot \frac{d_I}{d}. \tag{3}$$

With the fit value of $8 \cdot 10^4 \, s^{-1} \, at \, 93 \, °C$ (see Figure 2d), Eq.3 provides $\sigma \sim 1.5 \cdot 10^{-5} \, 1/(\Omega m)$. We note that it will slightly modify Eq.(2) as $f_o^{N_F} = \frac{P^2 + \gamma\sigma}{\gamma\varepsilon_o \cdot \varepsilon_I} \cdot \frac{d_I}{d}$. However, $\gamma\sigma \ll P^2$, i.e., the effect of electric conductivity can be neglected in the $N_F$ phase.

Importantly, $\varepsilon_A''$ measured at high frequencies for different cells show similar values regardless of the values of $d$ and $d_I$, Figure 3. This result confirms Eq.(1) of the PCG model, which for large frequencies, $\omega\tau_o \gg 1$, simplifies to $\varepsilon_A''(\omega) = \varepsilon_I \frac{d}{\omega\tau_o d_I}$. Since the relaxation time is $\tau_o = \frac{\varrho_{LC}\varepsilon_o\varepsilon_I d}{d_I}$, one finds

$$\varepsilon_A''\left(f \gg \frac{2\pi}{\tau_o}\right) \approx \frac{1}{2\pi f \varrho_{LC}\varepsilon_o}, \tag{4}$$

which does not depend on $d$ nor on $d_I$. To illustrate the overlap, Figure 3 shows together the $\varepsilon_A''(f)$ curves measured at different temperatures in the $N_F$ phase and for different $d_I$. We also plotted the fits for $d_I = 90 \, nm \, and \, 154 \, nm$. The best fits yield $\varrho_{LC}\varepsilon_o = (2.2 - 2.76) \cdot 10^{-8} \, s$, which corresponds to resistivity $\varrho_{LC} \sim (2.5 - 3.1) \cdot 10^3 \, \Omega m$. Assuming $\varrho_{LC} \sim \gamma/P^2$, one gets $\gamma \sim 5\text{-}10 \, Pa \cdot$ s, which is comparable to the viscosity values obtained from the $d_I/d$ dependence of the relaxation frequencies, as shown in Figure 2d.

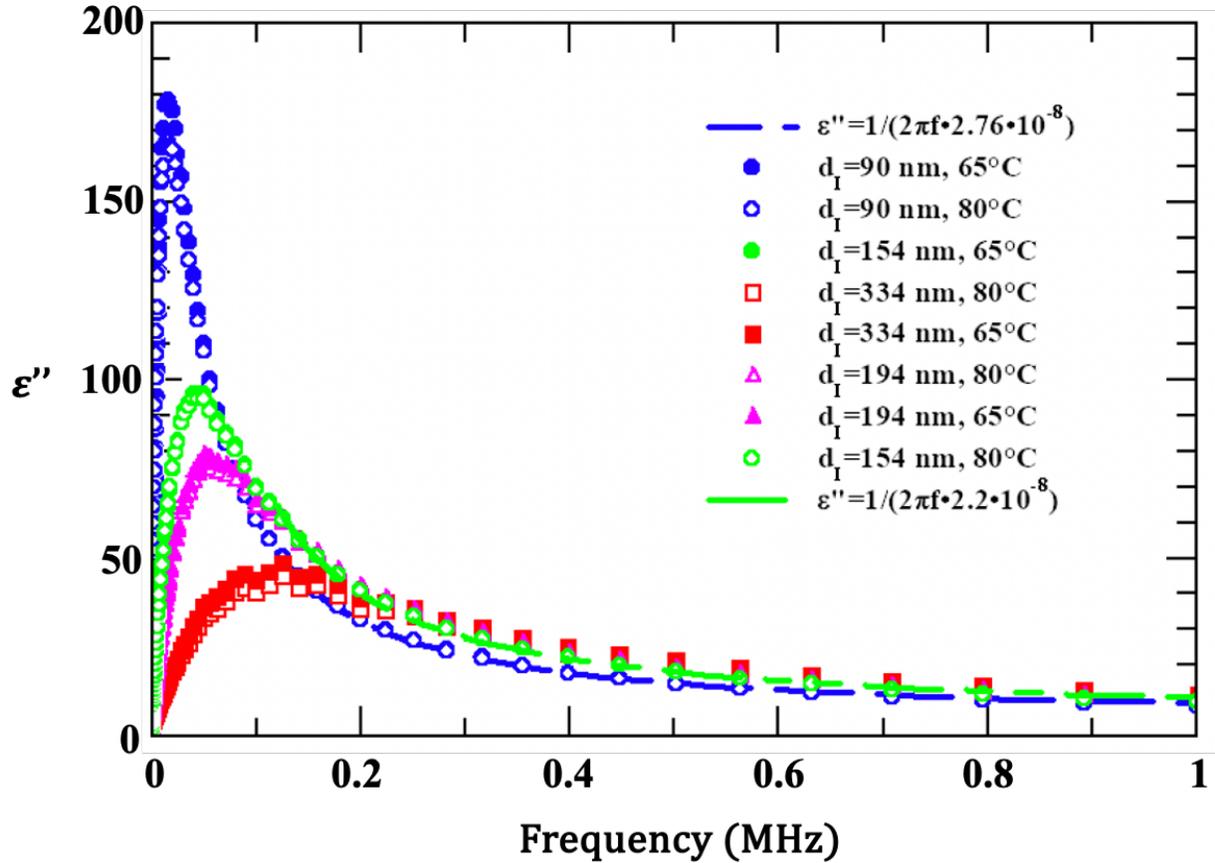

*Figure 3: The measured $\varepsilon_A''(f)$ curves in linear scale in the $N_F$ phase at 65 °C and 80 °C for all insulating layer thicknesses. Dashed blue and green lines correspond to fits to Eq.(4) for $d_I = 90\ nm$ and $154\ nm$, respectively.*

We have carried out additional measurements using bare ITO coated substrates. The frequency dependent spectra of $\varepsilon'$ and $\varepsilon''$ at selected temperatures are shown in Figure 4(a,b) for $d = 10.9\ \mu m$ and in Figure 4(c,d) for $d = 17.2\ \mu m$ films, respectively. One can see that in the isotropic and nematic phases $\varepsilon''$ decreases monotonously without showing any peak above $20\ Hz$. According to Eq.(3), the decreasing frequency of the peak position should be due the effectively much smaller insulating layer thickness on bare ITO substrates. For the same reason, in the $N_F$ phase the relaxation frequency on bare ITO substrates is also about one order of magnitude smaller than in the PI2555 coated cells. Importantly, the measured low frequency dielectric constants now approaching $\varepsilon_A(0) \sim 8 \cdot 10^3$ in the $d = 10.9\ \mu m$ sample and $\varepsilon_A(0) \sim 1.7 \cdot 10^4$ for $d = 17.2\ \mu m$ sample, similar to those reported by others in samples with Gold electrodes [3,4,6–9]. As the alignment even in bare ITO cells is close to planar, such as in the PI2555 coated cells, we conclude that the large difference is due to the much thinner insulating layer $d_I = \frac{\varepsilon_I}{\varepsilon_A(0)} d$, which assuming that $\varepsilon_I$ on the bare ITO is similar to dielectric polymers, provide $d_I \approx 4.8\ nm$ and $\approx 3.5\ nm$ for the 10.9 μm and 17.2 μm cells, respectively.

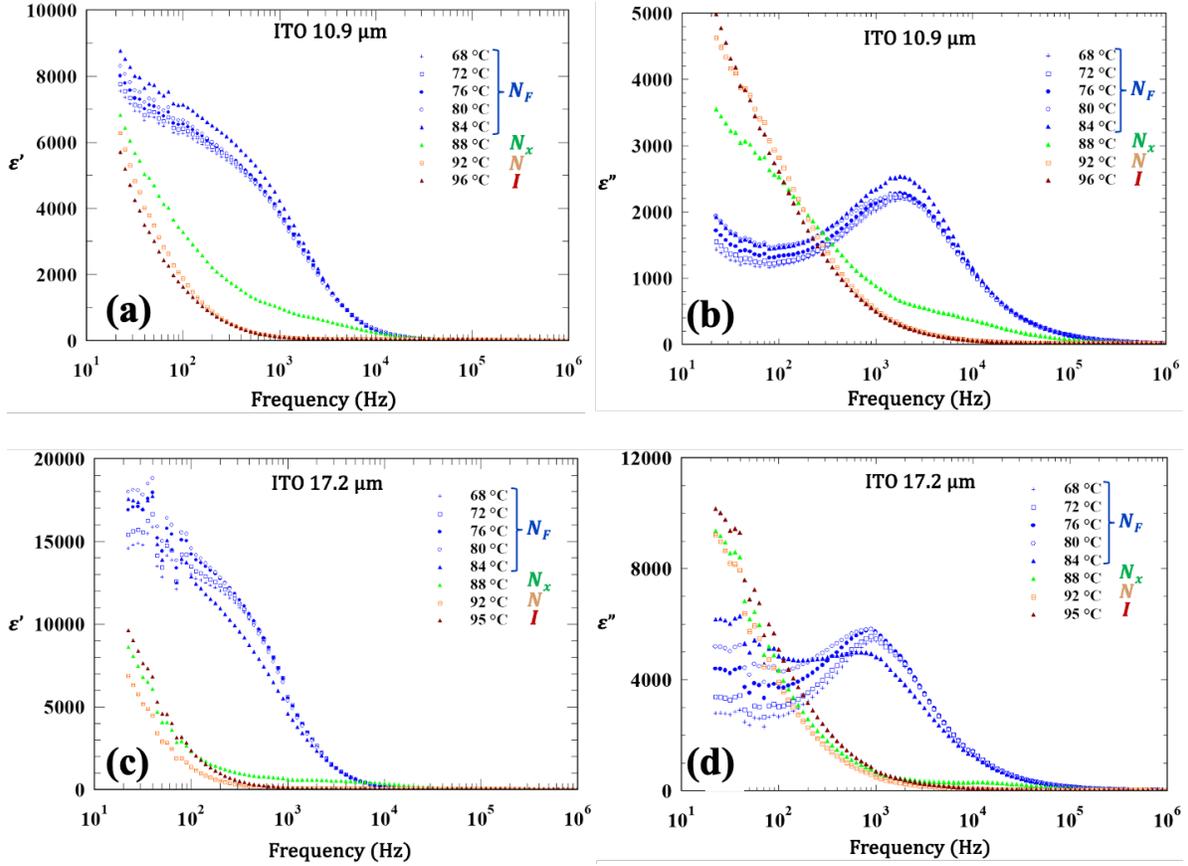

Figure 4: $\varepsilon'(f)$ and $\varepsilon''(f)$ of bare ITO coated sandwich cells at various temperatures. (a and b) : $\varepsilon'(f)$ and $\varepsilon''(f)$ of 10.9μm film, respectively; (c and d): $\varepsilon'(f)$ and $\varepsilon''(f)$ of 17.2 μm film, respectively.

This indicates that the thickness of the insulating layer on each substrate $\xi_I$ is $1.8\ nm < \xi_I < 2.4\ nm$, which is comparable to the molecular length. We note here that Clark et al. [11] state that even at conducting substrates (such as the bare ITO), there should be a layer of thickness

$$\xi_I = \sqrt{\frac{\varepsilon_o \varepsilon_{LC} K}{P^2}}, \qquad (5)$$

where the polarization space charge is expelled and should be considered as non-ferroelectric insulating layer. In the above expression $K$ is an effective Frank elastic constant and $\varepsilon_{LC}$ is the bare $N_F$ dielectric constant. Using $K \sim 10\ pN$, we can estimate that $\varepsilon_{LC} \approx \frac{\xi_I^2 P^2}{K \varepsilon_o} \sim 120$ for the 10.9 μm cell and $\varepsilon_{LC} \sim 230$ for 17.2 μm cell. The difference between the calculated apparent dielectric constants can be due to slight alignment differences, or simply due to the fact that the apparent thickness of the insulating layers might be different from $\xi_I$ because of oxidation or any dirt. For this reason, the above estimated $\varepsilon_{LC}$ values are likely larger than the actual values.

Using $d_I \approx 4.8\ nm$ and $d_I \approx 3.5\ nm$ for the 10.9 μm and the 17.2 μm cells, from Eq. (2) we can also estimate the rotational viscosity values to be $\gamma \approx 4.4\ Pas$ and $4.2\ Pas$, respectively. These values are similar that those we obtained for the PI coated cells, thus indicating that the alignment on the bare ITO coated cells is also very close to planar and represents rotations involving mainly splay.

To summarize, detailed analysis of our measurements on UUZU-4-N in the frame of the polarization-capacitance Goldstone (PCG) model of Clark et al [11] provided quantitative agreement in terms of the measured dielectric constant as a function of $d/d_I$, verifying that the measured dielectric constant is determined by the capacitance of the insulating layer and does not represent the dielectric permittivity of the ferroelectric nematic. The often reported extraordinary high values of the apparent dielectric permittivity of the $N_F$ phase are the result of "block polarization reorientation" which screens the applied electric field within the $N_F$ cell. The applied voltage acts only at the insulating layers such as polyimide coatings. The apparent dielectric permittivity $\varepsilon_A(0) = \varepsilon_I d/d_I$ is then a characteristic of the insulating layers. A large $d/d_I$ value produces an apparent $\varepsilon_A(0)$ on the order of $10^4$. Comparing the measured $d_I/d$ dependence of the relaxation frequency with Eq.(2), further confirms the PCG model. We could then use our results on films with bare ITO substrates to estimate the thickness of dielectric layer adjacent to the ITO substrate to be $\frac{d_I}{2} \sim 2$ nm. With this, using again the PCG model [11], we estimate the dielectric constant when the polarization is perpendicular to the electric field, to be $\varepsilon_\perp \sim 10^2$.

*Acknowledgement:* This work was financially supported by US National Science Foundation grant DMR-2210083 (A.J. and S.S.), DMR- 2215191 (O.D.L., digital holographic microscopy studies), and ECCS- 2122399 (O.D.L., analysis of switching). The material UUZU-4-N was provided by Merck Electronics KGaA, Darmstadt, Germany.

# Supplemental Material

# Dielectric properties of a ferroelectric nematic material: quantitative test of the polarization-capacitance Goldstone mode

Alex Adaka[1,3], Mojtaba Rajabi[2], Nilanthi Haputhantrige[2,3], S. Sprunt[2,3], O.D. Lavrentovich[1,2,3] and A. Jákli[2,3]

[1]Materials Science Graduate Program, Kent State University, Kent OH, 44242, USA

[2]Department of Physics, Kent State University, Kent OH, 44242, USA

[3]Advanced Materials and Liquid Crystal Institute, Kent State University, Kent OH, 44242, USA

1. Polarization and Threshold Voltage Measurements

Polarization measurements were carried out in 10 μm thick planar aligned films by applying 80 Hz triangular voltage of amplitudes $U_s$ (Figure S1) between in-plane ITO strips separated by a distance $L = 0.5$ mm. The set-up is similar to the one described by Chen et al [1]. The temperature dependence of the calculated ferroelectric polarization is plotted against the left axis of Figure S1. The highest value of P is slightly larger than that of RM734 [1] and DIO [2] likely due to the slightly larger molecular dipole ($\mu \approx 11.9$ Debye) of UUZU-4-N. The measured values can be compared to a theoretical value $P = |\vec{P}| = \frac{|\langle\vec{\mu}\rangle|}{V_m} = S_P \cdot \frac{\mu \cdot \varrho \cdot N_A}{M}$, where $S_P$ is the polar order parameter, and $V_m$ is the volume occupied by a molecule that can be related to the mass density $\rho \approx 1.3 \frac{g}{cm^3}$, molar mass $M \approx 463$ g/mol, and the Avogadro's number $N_A \approx 6 \cdot 10^{23}$/mol. Assuming perfect polar order, $S_P = 1$, we get $P = 6.7 \cdot 10^{-2}$ C/m$^2$, which is smaller than the measured highest value of $P = 7.5 \cdot 10^{-2}$ C/m$^2$. As $S_P$ should be smaller than 1, the measured values are likely affected by ionic contributions. Assuming $S_P = 0.95$ that was found by the Boulder group for RM734 [1], we estimate that the maximum polarization is only $P = 6.3 \cdot 10^{-2}$ C/m$^2$.

The threshold voltage $U_s$ needed for full polarization switching is plotted on the right axis. While on cooling in the $N_F$ phase $U_s$ increases slightly from ∼10 V to ∼15 V in 20 °C temperature range, in the $N_X$ phase the switching voltage drops from 130 V to 10 V within a 6 °C temperature range, while the polarization value sharply increases. This may indicate that the $N_X$ phase is either a $SmZ_A$ [3] or a splayed nematic ($N_S$) [4] phase, or something else. Its nature will be the subject of further research.

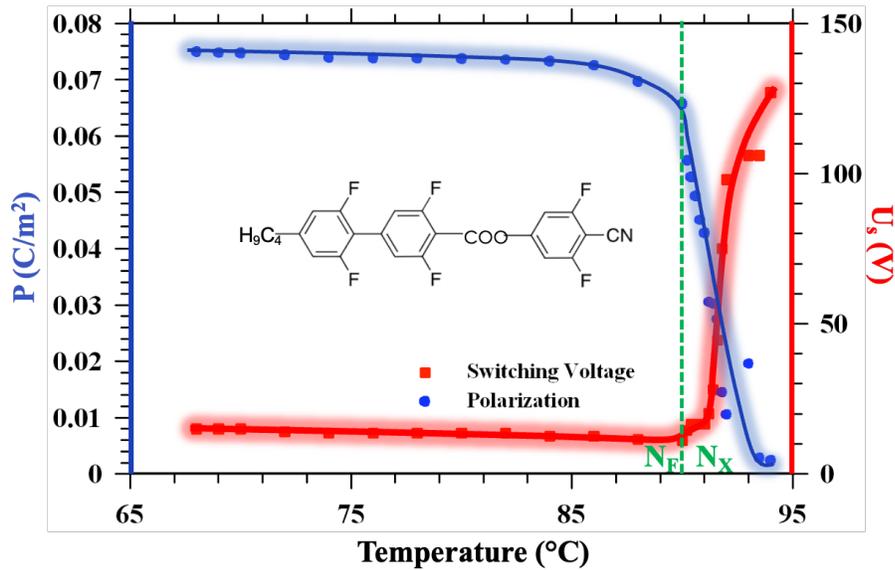

Figure S1: Temperature dependence of the ferroelectric polarization (left axis) and of the voltage $U_s$ (right axis) needed for full polarization switching. The width of the shading indicates the error of the measurements. Inset shows the molecular structure of the studied UUZU-4-N material.

2. Switching Time Measurements

The voltage dependence of the switching time is shown for several temperatures in Figure S2. The rotational viscosity γ was estimated from the voltage dependence of the switching time $\tau \approx \frac{\gamma \cdot L}{|P|} \cdot \frac{1}{V}$, which is valid if the effect of the dielectric coupling and the surface anchoring can be neglected compared to the ferroelectric coupling. [5] The best fit to the τ(V) curve in Figure S2 gives $0.96\ Pa \cdot s$ and $0.85\ Pa \cdot s$ viscosity values at 75°C and 85°C, respectively.

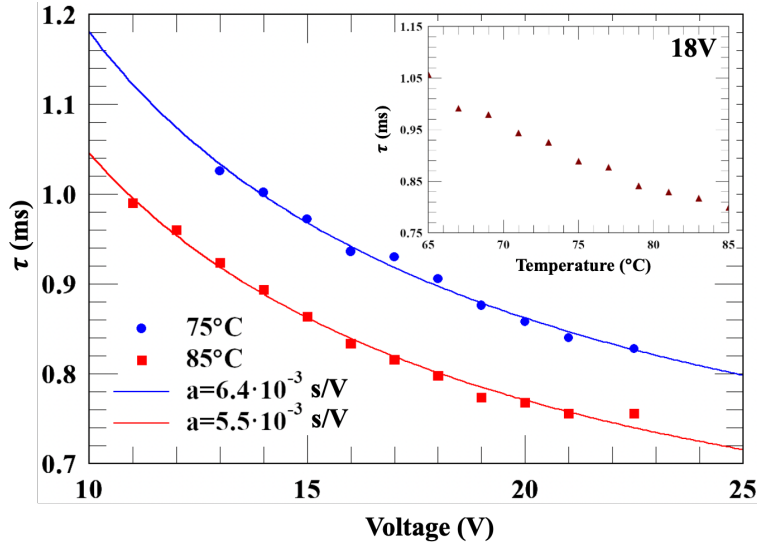

Figure S2: Voltage dependence of the switching time $\tau$ at 75°C and 85°C. The rotational viscosity $\gamma$ is calculated from the fit values $a = \frac{\gamma \cdot L}{P}$, where L=0.5 mm is the gap between in-plane electrodes and P is the spontaneous polarization shown in Figure S1. The inset shows the temperature dependence of the switching time.

3. **Sample Preparation**

The cells represent an $N_F$ slab of thickness $d = (10.5 - 13.1)$ μm $\pm$ 0.2 μm confined between two glass plates with transparent Indium Tin Oxide (ITO) electrodes. The electrodes were spin-coated with the polyimide PI2555 solutions in T-9039 (HD Microsystems) using a spin coater WS-650MZ-23NPP (Laurell Technologies), which was spun for 1 s @ 500 rpm, 30 s @ 1500 rpm and 1 s @ 50 rpm. The coated substrates were baked for 5 mins @ 80°C for 1 hour @ 275°C and buffed to provide a uniform alignment of the $N_F$. To explore the effect of the insulating layer thickness on the dielectric permittivity measurements, we used four different PI2555/solvent ratios, namely, 1:9, 1:7, 1:5, and 1:3, which produced four different thicknesses $d_I/2$ of the buffed PI2555 alignment layer: $45 \pm 5$ nm, $77 \pm 5$ nm, $97 \pm 5$ nm and $167 \pm 5$ nm, respectively. The thicknesses $d_I/2$ of the alignment layers were measured by a digital holographic microscope (Lyncée Tec), with vertical resolution < 1 nm. We scratched the PI2555 - coated glasses with a sharp blade to make 4 grooves, then we measured the layer thickness at 5 different places along the grooves and averaged the values. The $\pm$ 5 nm error is due to the inhomogeneity of the coated PI2555 film and of the scratching. Within the measuring errors, the obtained values are in good agreement with ellipsometry measurement. The two PI2555 coated substrates together formed four different cells with $d_I = 90\,\text{nm}$, $154\,\text{nm}$, $194\,\text{nm}$ and $334\,\text{nm}$. In addition to the cells with different preset thicknesses $d_I$ of the insulating layers, we explored cells with bare ITO electrodes and no PI2555 coatings.

The liquid crystal film thickness d was measured for empty cells by the interference technique using OceanOptics 2000 spectrophotometer. Each cell was filled with the LC in the isotropic phase at 110°C by capillary action and each measurement was carried out at fixed temperatures on cooling.

4. **Dielectric spectroscopy**

The dielectric spectra have been measured using an HP 4284A impedance analyzer at frequencies between 20 Hz and 1 MHz . The real $\varepsilon'$ and imaginary $\varepsilon''$ components of the complex dielectric permittivity were calculated assuming an equivalent parallel $R - C$ circuit where the capacitance $C_P$ and resistance $R_P$ values were recorded by the Impedance analyzer. Neglecting any resistance of the electrode and wires $r_e < 1\,\text{k}\Omega$ (this is a good approximation up to 1 MHz), these $C_P$ and $R_P$ values give $\frac{1}{Z_S} = \frac{1}{R_P} + i\omega C_P$, where $Z_S$ is the impedance of the film. The real and imaginary part of the dielectric constant can be calculated as $\varepsilon' = \frac{C_P}{C_o}$ and $\varepsilon'' = \frac{1}{C_o \omega R_P}$, respectively, where $C_o = \varepsilon_o \frac{A}{d}$ is the capacitance of the empty cell with area A that can be obtained from the low frequency limit of the impedance of the empty cell measured before filling with the liquid crystal [6]. In our measurements we have used the values calculated from the cell area and thickness. We also compared with the measured empty cell data and found good agreement with the measured value (error was always less than 10%).

5. **References**

[1] X. Chen et al., First-Principles Experimental Demonstration of Ferroelectricity in a Thermotropic Nematic Liquid Crystal: Polar Domains and Striking Electro-Optics, Proc Natl Acad Sci USA **117**, 14021 (2020).